\documentstyle[twocolumn,seceq,epsbox]{jpsj}

\def\runtitle{Susceptibility, Magnetization Process and ESR Studies on the Helical Spin System RbCuCl$_{3}$}
\def\runauthor{Syuji {\sc Maruyama} {\it et. al.}}

\title
{
Susceptibility, Magnetization Process and ESR Studies \\ on the Helical Spin System RbCuCl$_{3}$ 
}

\author
{ 
Syuji {\sc Maruyama}\footnote{E-mail: syuji@lee.phys.titech.ac.jp.}, Hidekazu {\sc Tanaka}\footnote{E-mail: tanaka@lee.phys.titech.ac.jp.}, Yasuo {\sc Narumi}$^{1,2}$, Koichi {\sc Kindo}$^{2,1}$, \\
Hiroyuki {\sc Nojiri}$^{3,1}$, Mitsuhiro {\sc Motokawa}$^{3,1}$ and Kazukiyo {\sc Nagata}$^4$
}

\inst
{
Department of Physics, Tokyo Institute of Technology, Oh-okayama, Meguro-ku, Tokyo 152-8551\\
$^1$CREST, Japan Science and Technology, Kawaguchi, Saitama 332-0012\\
$^2$Research Center for Materials Science at Extreme Conditions, Osaka University, Toyonaka,\\ 
Osaka 560-0043\\
$^3$Institute for Material Research, Tohoku University, Aoba-ku, Sendai 980-8577\\
$^4$Department of Physics, Faculty of Engineering, Kanagawa University, Rokkakubashi, Kanagawa-ku, Yokohama 221-8686
}

\recdate
{
\hspace{7cm}}

\abst
{
The static and dynamic magnetic properties of an $S=1/2$ stacked triangular antiferromagnet RbCuCl$_{3}$ with a helical spin structure due to lattice distortion were investigated by magnetic susceptibility, high-field magnetization process and ESR measurements. The magnetic susceptibilities were analyzed by high-temperature expansion approximation in terms of ferromagnetic intrachain coupling and the antiferromagnetic interchain coupling. The magnetization saturates at $H_{\rm s}\approx 66.8$ T at 1.5 K, the value of which is twice that of CsCuCl$_3$. A small magnetization jump indicative of a phase transition was observed at $H_{\rm c}=21.2$ T $\approx (1/3)H_{\rm s}$ for $H{\perp}c$. ESR modes observed for $H\parallel c$ are well described by the calculation based on the helical spin structure. From the present measurements, the ferromagnetic intrachain and two kinds of antiferromagnetic interchain exchange interactions, and the planar anisotropy energy were determined as $J_0/k_{\rm B}=25.7$ K, $J_1/k_{\rm B}=-10.6$ K, $J_1'/k_{\rm B}=-17.4$ K, and ${\Delta}J_0/k_{\rm B}=-0.45$ K, respectively.
}

\kword
{
RbCuCl$_{3}$, distorted triangular antiferromagnet, helical spin structure, susceptibility, high-field magnetization process, field-induced phase transition, ESR
}

\begin{document}
\sloppy
\maketitle

\section{Introduction}
\begin{figure}[b]
\begin{center}
  \epsfigure{file=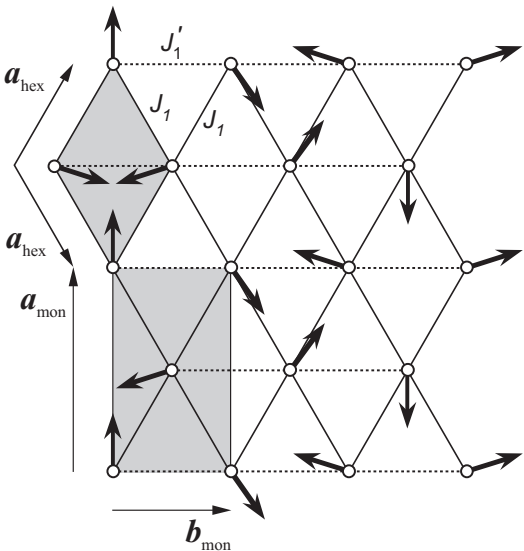,width=75mm}
\end{center}
\caption{Interchain exchange interactions, $J_1$ (solid lines) and $J'_1$ (dotted lines), and incommensurate spin structure in the basal plane. Shaded rectangles and diamonds are the unit cells for monoclinic and hexagonal representations, respectively.}
\label{fig:1}
\end{figure}
Triangular antiferromagnets of the hexagonal ABX$_3$ type with the CsNiCl$_3$ structure exhibit a rich variety of magnetic phase transitions due to spin frustration and quantum fluctuation.
\cite{Collins} Recently, the magnetic properties of CsCuCl$_3$ have been actively investigated. CsCuCl$_3$ exhibits field-induced phase transition for $H\parallel c$, which is caused by the competition between planar anisotropy and quantum fluctuation.\cite{Nojiri1,Nikuni1,Schotte} A small amount of Co$^{2+}$ doped in CsCuCl$_3$ produces successive magnetic phase transitions,\cite{Ono1} and the low-temperature phase was found to be an oblique triangular antiferromagnetic phase in which spins form triangular structure in a plane tilted from the basal plane.\cite{Ono2} \par
The magnetic properties of the closely related RbCuCl$_3$ have been less extensively studied. At high temperatures, RbCuCl$_3$ exhibits the highly symmetric CsNiCl$_3$ structure (space group $P6_3/mmc$). With decreasing temperature, RbCuCl$_3$ transforms into an orthorhombic structure ($Pbcn$) at $T_{\rm t1}=340$ K and further into a monoclinic structure ($C2$) at $T_{\rm t2}=260$ K due to the Jahn-Teller effect.\cite{Crama,Harada} RbCuCl$_3$ undergoes magnetic phase transition at $T_N\approx 19$ K.\cite{Tazuke1} The magnetic susceptibility of RbCuCl$_3$ was first investigated by Tazuke {\it et al}.\cite{Tazuke1} 
They analyzed their susceptibility data in terms of antiferromagnetic intrachain coupling and ferromagnetic interchain coupling. Recent neutron powder diffraction by Reehuis {\it et al}.\cite{Reehuis} revealed that an incommensurate helical spin structure which is close to the 120$^{\circ}$ spin structure is realized in the basal plane as shown in Fig. 1, and that spins are arranged ferromagnetically along the $c$-axis. The spin structure is expressed by the ordering vector $\mib Q_0=(0, 0.5985, 0)$ for the monoclinic representation and $\mib Q_0=(0.2993, 0.2993, 0)$ for the hexagonal representation.\cite{Reehuis} This fact indicates that the intrachain interaction is ferromagnetic and the interchain interaction is antiferromagnetic. \par 
The incommensurate helical spin structure of RbCuCl$_3$ arises from the breaking of the hexagonal symmetry as observed in RbMnBr$_3$,\cite{Glinka,Kato,Perez1} {\it i.e.}, the interchain interactions $J_1$ and $J'_1$ are not equivalent, as shown in Fig. 1. \par
The magnetic interactions in RbCuCl$_3$ may be written by 
\begin{eqnarray}
{\cal H}=&-&\sum_{l,j}2J_0({\mib S}_{l,j}\,\cdot\,{\mib S}_{l,j+1})
-\sum_{l,j}2{\Delta}J_0(S_{l,j}^z\,S_{l,j+1}^z)  \nonumber \\
&-&\sum_{l,m,j}2J_1^{lm}{\mib S}_{l,j}\,\cdot\,{\mib S}_{m,j} ,
\end{eqnarray}
where ${\mib S}_{l,j}$ is the spin-$\frac{1}{2}$ operator on the $j$-th Cu$^{2+}$ ion in the $l$-th chain, $J_0$ is the ferromagnetic intrachain interaction, ${\Delta}J_0$ is the anisotropy of the easy-plane type $({\Delta}J_0<0)$, and $J_1^{lm}$ is the antiferromagnetic interchain interaction, $J_1^{lm}=J_1$ along the $[1, 0, 0]$ and $[0, 1, 0]$ directions, and $J_1^{lm}=J_1'$ along the $[1, 1, 0]$ direction in the hexagonal representation. In this paper, we use the hexagonal representation for the reciprocal lattice vector $\mib Q$. The ordering vector $\mib Q_0=(Q_0, Q_0, 0)$ is given by
\begin{eqnarray}
\cos (2{\pi}Q_0)=-\frac{J_1}{2J'_1} .
\end{eqnarray}
Substituting $Q_0=0.2993$ into eq. (1.2), we have $J_1/J'_1=0.61$. The origin of the incommensurate helical spin structure in RbCuCl$_3$ is different from that for CsCuCl$_3$, in which the competition between the ferromagnetic intrachain interaction and the antisymmetric interaction of the Dzyaloshinsky-Moriya type gives rise to the incommensurate helical spin structure along the $c$-axis.\cite{Adachi} \par
Recent precise specific heat measurements by P\'{e}rez-Willard {\it et al.}\cite{Perez2} revealed that RbCuCl$_3$ exhibits two magnetic phase transitions at $T_{\rm N1}=18.91$ K and $T_{\rm N2}=18.87$ K at zero field. These two transition temperatures are very close as observed in RbMnBr$_3$.\cite{Perez1} It is considered that the weak in-plane anisotropy produces the split of the phase transition, because the successive transitions cannot be explained within the Hamiltonian of eq. (1.1). For $H\perp c$, the temperature range of the intermediate phase between $T_{\rm N1}$ and $T_{\rm N2}$ increses with increasing external field.\cite{Perez2} \par 
In order to investigate the static and dynamic magnetic properties of RbCuCl$_3$, we performed susceptibility, high-field magnetization and high-field ESR measurements using single crystals. Within the present measurements, we can determine all of the interactions in eq. (1.1). This paper is organized as follows. The experimental procedures are described in \S 2. The experimental results, their analyses and a discussion are given in \S 3. The final section is devoted to the conclusions.  \par
\begin{fullfigure}[t]
  \begin{minipage}{90mm}
    \begin{center}
      \epsfigure{file=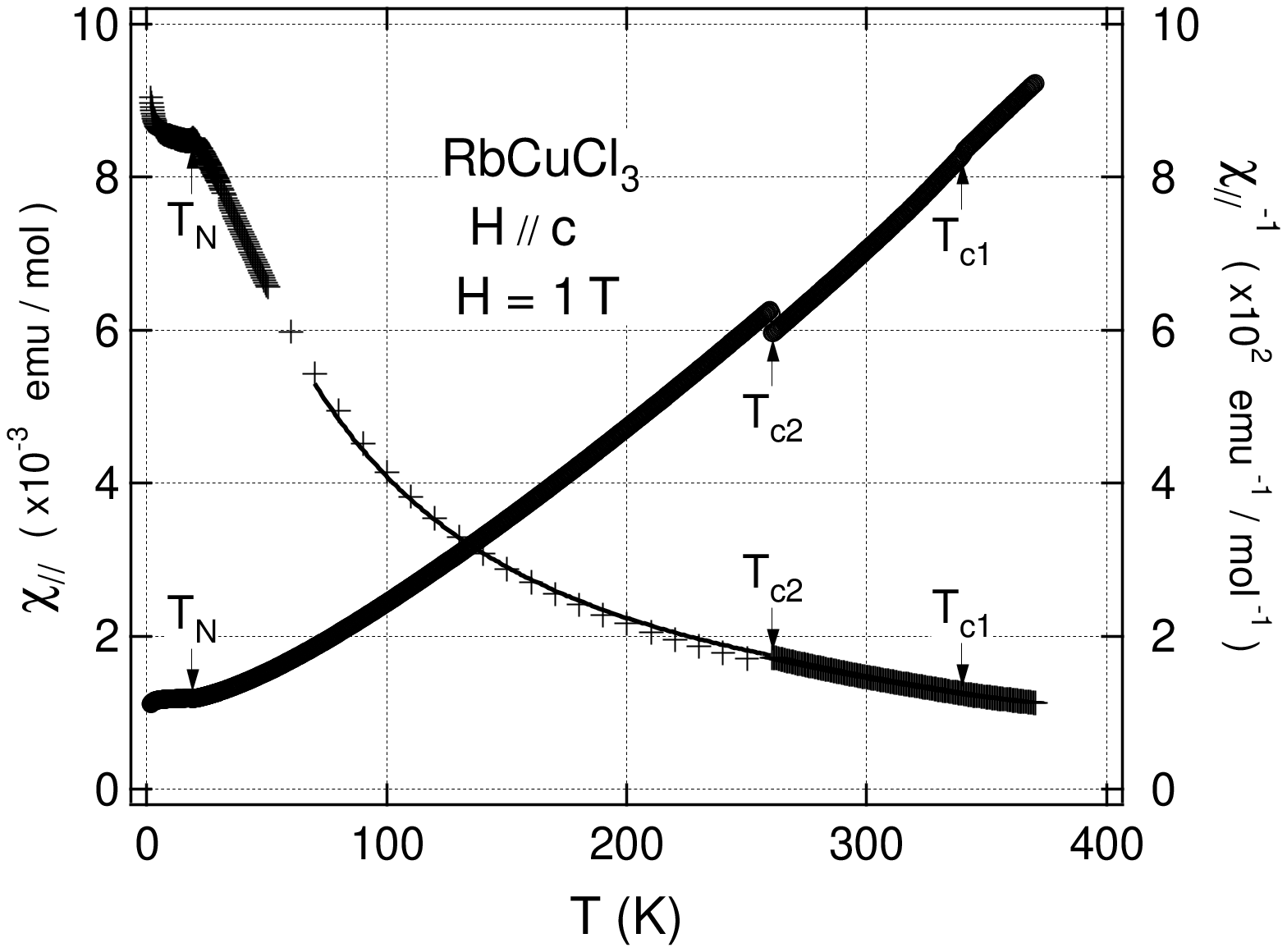,height=60mm}
      \vspace{2mm}
        (a)
    \end{center}
  \end{minipage}
  \begin{minipage}{90mm}
    \begin{center}
      \epsfigure{file=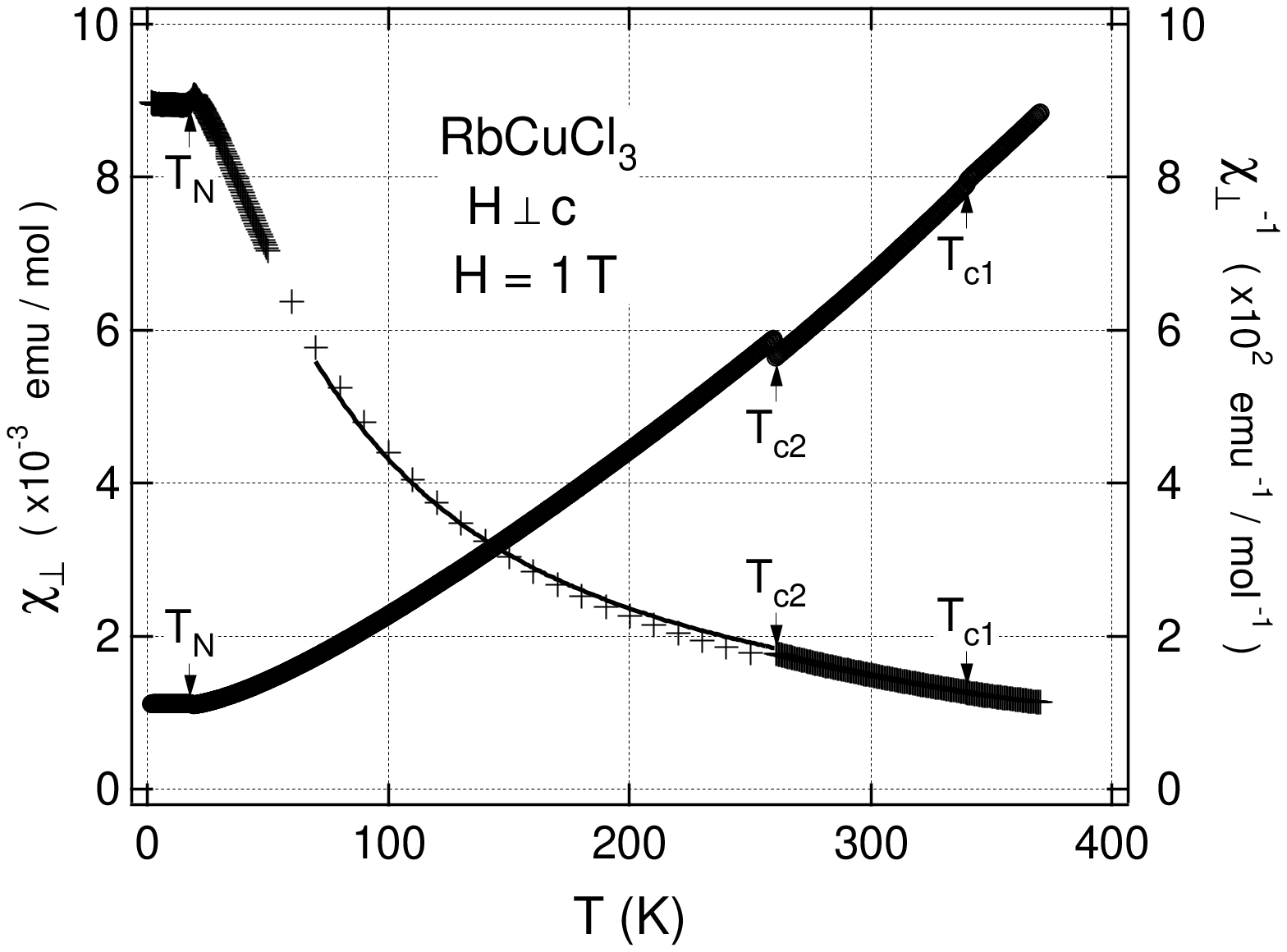,height=60mm}
      \vspace{2mm}
        (b)
    \end{center}
  \end{minipage}
\caption{Temperature dependence of the magnetic susceptibilities $\chi$ and inverse susceptibilities 
${\chi}^{-1}$ of RbCuCl$_{3}$ measured at $H=1$ T for (a) $H{\parallel}c$ and (b) $H{\perp}c$. Thin solid lines denote the fitting for $\chi$ by eq. (3.4). The data points for $\chi$ vs. $T$ were thinned out between 50 K and 260 K so that the fitting curves can be visible.}
\label{fig:2}
\end{fullfigure}

\section{Experimental Details}
Single crystals of RbCuCl$_3$ were grown by the vertical Bridgman method from the melt of an equimolar mixture of RbCl and CuCl$_2$ sealed in an evacuated quartz tube. The details of sample preparation are given in ref. 10. Since the crystals are hygroscopic, we treated them in a glove box filled with dry nitrogen. \par
The susceptibilities were measured between 1.8 K and 400 K at $H=1$ T using a SQUID magnetometer (Quantum Design MPMS XL). High-field magnetization measurement was performed at the Research Center for Materials Science at Extreme Conditions, Osaka University using an induction method with a multilayer pulse magnet which produces magnetic fields up to 56 T. ESR measurements at X ($\sim 9$ GHz) and K ($\sim 24$ GHz) bands frequencies were performed using a conventional reflection spectrometer with 80 kHz field modulation. The high-frequency, high-field ESR measurement was performed at the Institute for Material Research, Tohoku University using a multilayer pulse magnet which produces magnetic fields up to 30 T. FIR lasers ($323\sim762$ GHz), backward traveling wave tubes ($200\sim240$ GHz) and Gunn oscillators ($95\sim190$ GHz) were used as light sources. The Faraday configuration was taken in the present measurement. \par

\section{Results and Discussions}
\subsection{Susceptibility and Magnetization Curve}
Figure 2 shows the susceptibilities, $\chi_{\parallel}$ for $H\parallel c$ and $\chi_{\perp}$ for $H\perp c$, and their inverses as a function temperature. For $\chi$ vs. $T$, we thinned out the data points between 50 K and 260 K so that the fitting curves can be visible. The data were collected on heating. The susceptibilities were corrected for the Van Vleck paramagnetism, $\chi_{{\rm vv}\parallel}=0.68\,\times\,10^{-4}$ emu/mol and $\chi_{{\rm vv}\perp}=0.55\,\times\,10^{-4}$ emu/mol,\cite{Tazuke1} and the core diamagnetism $\chi_{{\rm dia}}=-1.09\,\times\,10^{-4}$ emu/mol.\cite{Selwood} Anomalies due to structural phase transitions were seen at $T_{\rm t1}=340$ K and $T_{\rm t2}=260$ K. The magnetic phase transition was observed at $T_{\rm N}=19$ K, as shown in Fig. 3. Within our experimental resolution, we could not distinguish the two phase transitions. \par 
\begin{figure}[htbp]
  \begin{center}
    \epsfigure{file=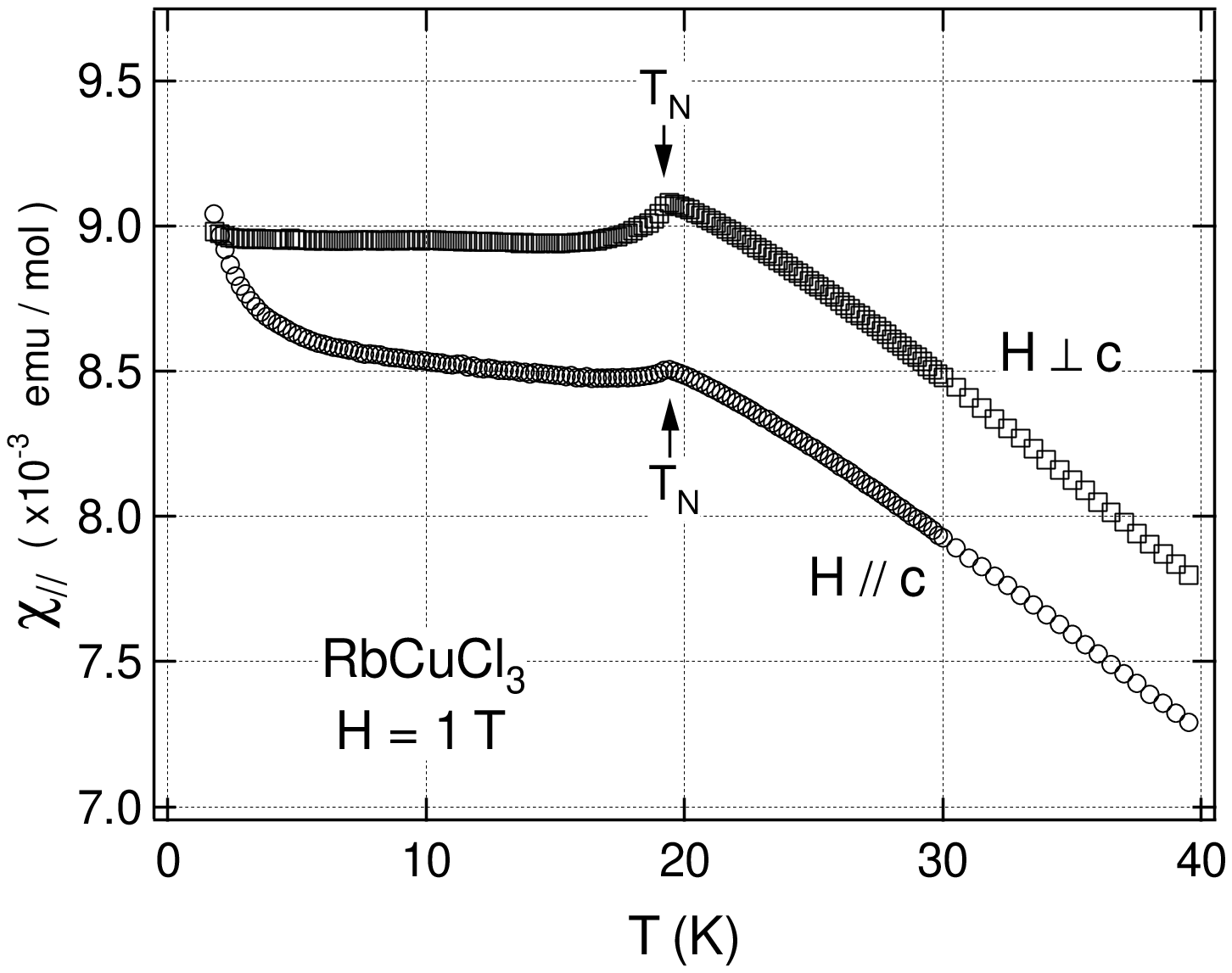,width=80mm}
  \end{center}
  \caption{Low-temperature susceptibilities for RbCuCl$_{3}$. N\'{e}el temperatures are indicated by arrows.}
  \label{fig:3}
\end{figure}
Figure 4 shows the magnetization curves measured at $T=1.3$ K for $H\parallel c$ and $H\perp c$. Unfortunately, the saturation cannot be achieved by magnetic fields up to 56 T. However, since the $g$-factor was obtained by the present ESR measurements as $g_{\parallel}=2.13$ and $g_{\perp}=2.14$ for $H\parallel c$ and $H\perp c$, respectively, the saturation magnetization could be estimated as $M_{\rm s}=1.06\ \mu_{\rm B}$ for $H\parallel c$ and $M_{\rm s}=1.07\ \mu_{\rm B}$ for $H\perp c$. Extrapolating the magnetization curve up to the saturation magnetization, we evaluate the saturation field as $H_{\rm s}=66.8$ T for $H\parallel c$ and $H_{\rm s}=66.4$ T for $H\perp c$. As shown in Fig. 4(b), a magnetization jump indicative of a phase transition is observed at $H_{\rm c}=21.2$ T $\approx (1/3)H_{\rm s}$ for $H\perp c$, while for $H\parallel c$, no anomaly is seen. \par
\begin{fullfigure}[htbp]
  \begin{minipage}{90mm}
    \begin{center}
      \epsfigure{file=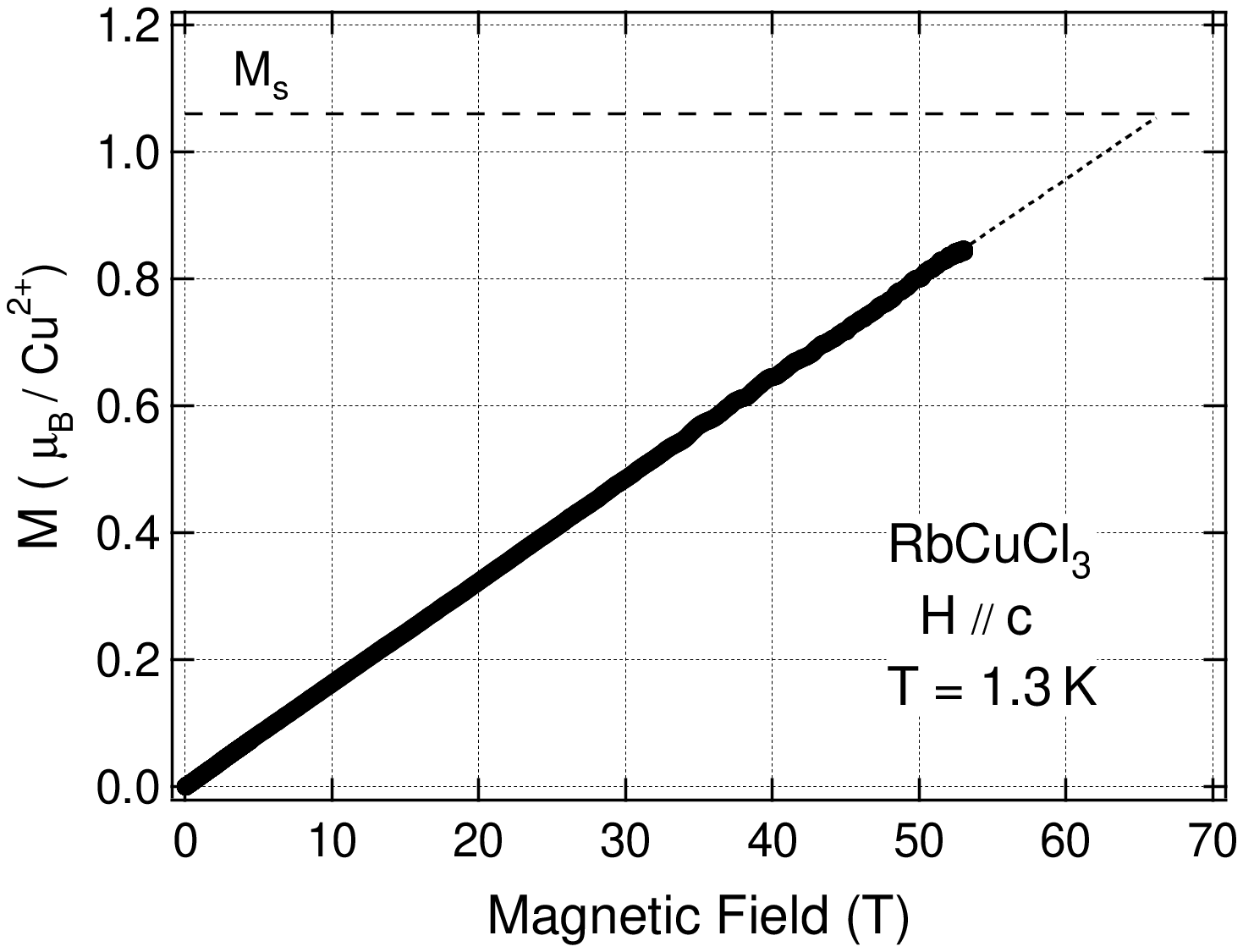,height=60mm}
         (a)
    \end{center}
  \end{minipage}
  \begin{minipage}{90mm}
    \begin{center}
      \epsfigure{file=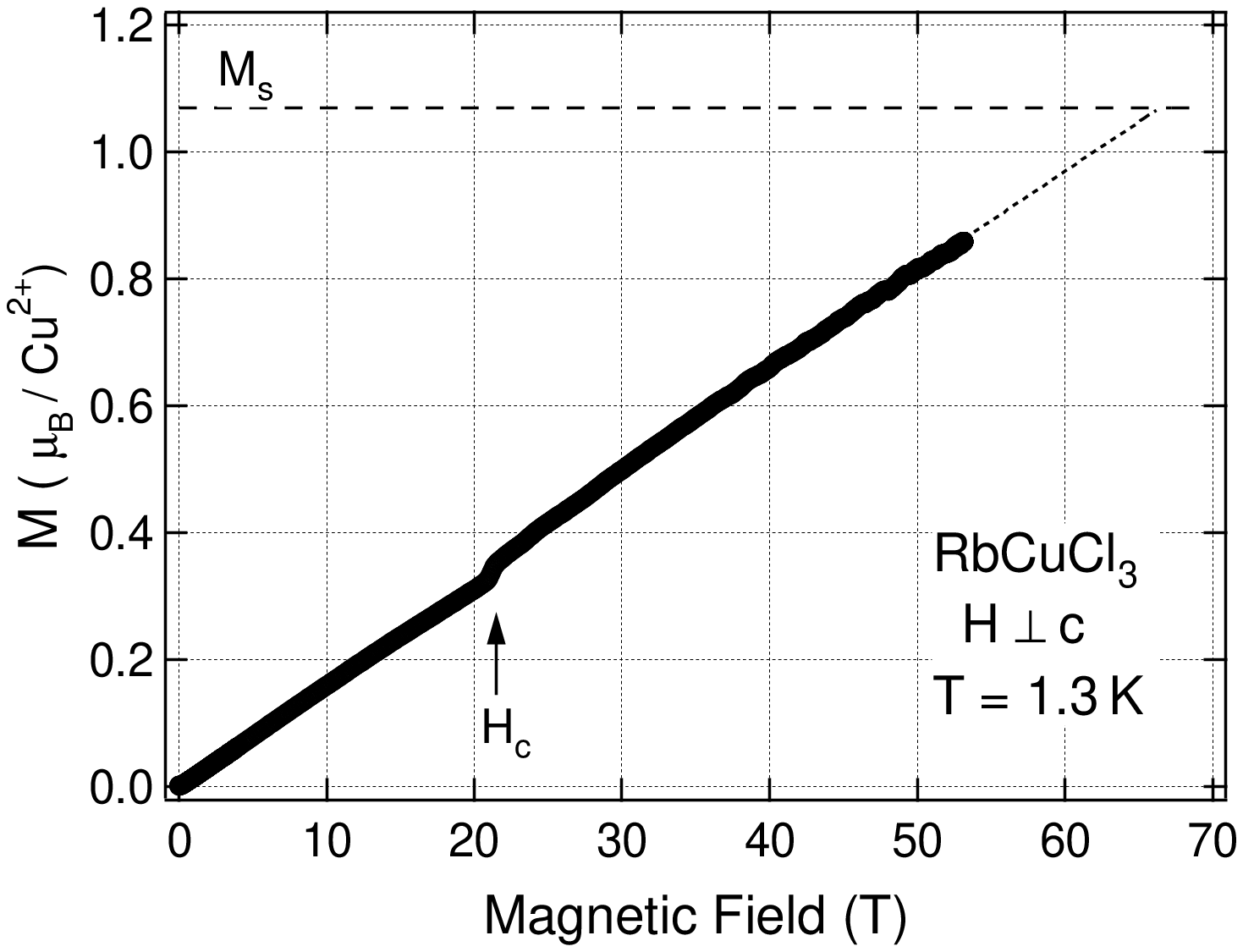,height=60mm}
        (b)
    \end{center}
  \end{minipage}
  \caption{Magnetization curves of RbCuCl$_{3}$ measured at 1.5 K for (a) $H{\parallel}c$ and (b) $H{\perp}c$.   Dotted lines are the extrapolations of the magnetization curves. Dashed lines are the saturation magnetization   expected from the $g$-factors.}
  \label{fig:4}
\end{fullfigure}
According to a recent neutron scattering experiment,\cite{Reehuis} spins lie in the basal plane and form a incommensurate helical structure which is close to the 120$^{\circ}$ spin structure. When an external field is applied along the $c$-axis, spins are raised from the basal plane, so that an umbrella structure is realized. The angle $\phi$ between the spin and the basal plane is given by
\begin{eqnarray}
\sin{\phi}=\frac{g_{\parallel}{\mu}_{\rm B}H}{2\left\{J({\mib Q}_0)-J(0)\right\}S} ,
\end{eqnarray}
where $J(\mib Q)$ is the Fourier transform of the exchange interactions and is given by
\begin{eqnarray}
J(\mib Q)=2J_0+4J_1\cos (2{\pi}Q)+2J_1'\cos (4{\pi}Q) ,
\end{eqnarray}
for $\mib Q=(Q, Q, 0)$. In eq. (3.1), we neglected the planar anisotropy ${\Delta}J_0$ which is much smaller than the exchange interactions, as we will see in the next subsection. Since the saturation of the magnetization occurs at $\phi=90^{\circ}$, the saturation field $H_{\rm s}$ for $H\parallel c$ is expressed as
\begin{eqnarray}
g_{\parallel}{\mu}_{\rm B}H_{\rm s}&=&2\{J({\mib Q}_0)-J(0)\}S \nonumber \\
&=&4J_1\{\cos (2{\pi}Q_0)-1\}\nonumber \\
& &{}\hspace{5mm}+2J'_1\{\cos (4{\pi}Q_0)-1\} .
\end{eqnarray}
Substituting $H_{\rm s}=66.8$ T, $g_{\parallel}=2.13$, $Q_0=0.2993$ and $J_1/J'_1=0.61$ into eq. (3.3), we obtain $J_1/k_{\rm B}=-10.6$ K and $J_1'/k_{\rm B}=-17.4$ K. \par
Next, we will evaluate the intrachain exchange interaction $J_0$ using the susceptibility data and the values of $J_1$ and $J'_1$. The susceptibility is expressed by the high-temperature expansion up to the third order as \cite{Tazuke1,Rushbrooke1,Rushbrooke2,Tazuke2} 
\begin{eqnarray}
\chi=\frac{Ng^2{\mu}_{\rm B}^2}{4k_{\rm B}T}\left\{1+\sum_{n=1}^3a_n\left(\frac{J_0}{k_{\rm B}T}\right)^n\right\} ,
\end{eqnarray}
with
\begin{eqnarray}
a_1=1+3R , \qquad a_2=6R(1+R) , \nonumber \\
\qquad a_3=-(1/3)+3R+21R^2+8.5R^3 . 
\end{eqnarray}
Here, $N$ is the number of spins and $R={\bar J}_1/J_0$, where ${\bar J}_1=(2J_1+J'_1)/3$ is the average of interchain exchange interactions. The value of ${\bar J}_1$ is ${\bar J}_1/k_{\rm B}=-12.9$ K. We fix the value of ${\bar J}_1$ and fit eq. (3.4) to the susceptibilities $\chi_{\parallel}$ and $\chi_{\perp}$ over the range from 70 K to 260 K. Then, we obtain $J_0/k_{\rm B}=25.7$ K, $g_{\parallel}=2.24$ and $g_{\perp}=2.30$. These $g$-factors are somewhat larger than $g_{\parallel}=2.13$ and $g_{\perp}=2.14$ obtained by the present ESR measurements. The value of $J_0$ is very close to $J_0/k_{\rm B}=25.1$ K reported by Reehuis {\it et al}.\cite{Reehuis} Solid lines in Fig. 2 are fitting curves of eq. (3.4). \par
\subsection{ESR}
\begin{fullfigure}[tbp]
  \begin{minipage}{90mm}
    \begin{center}
      \epsfigure{file=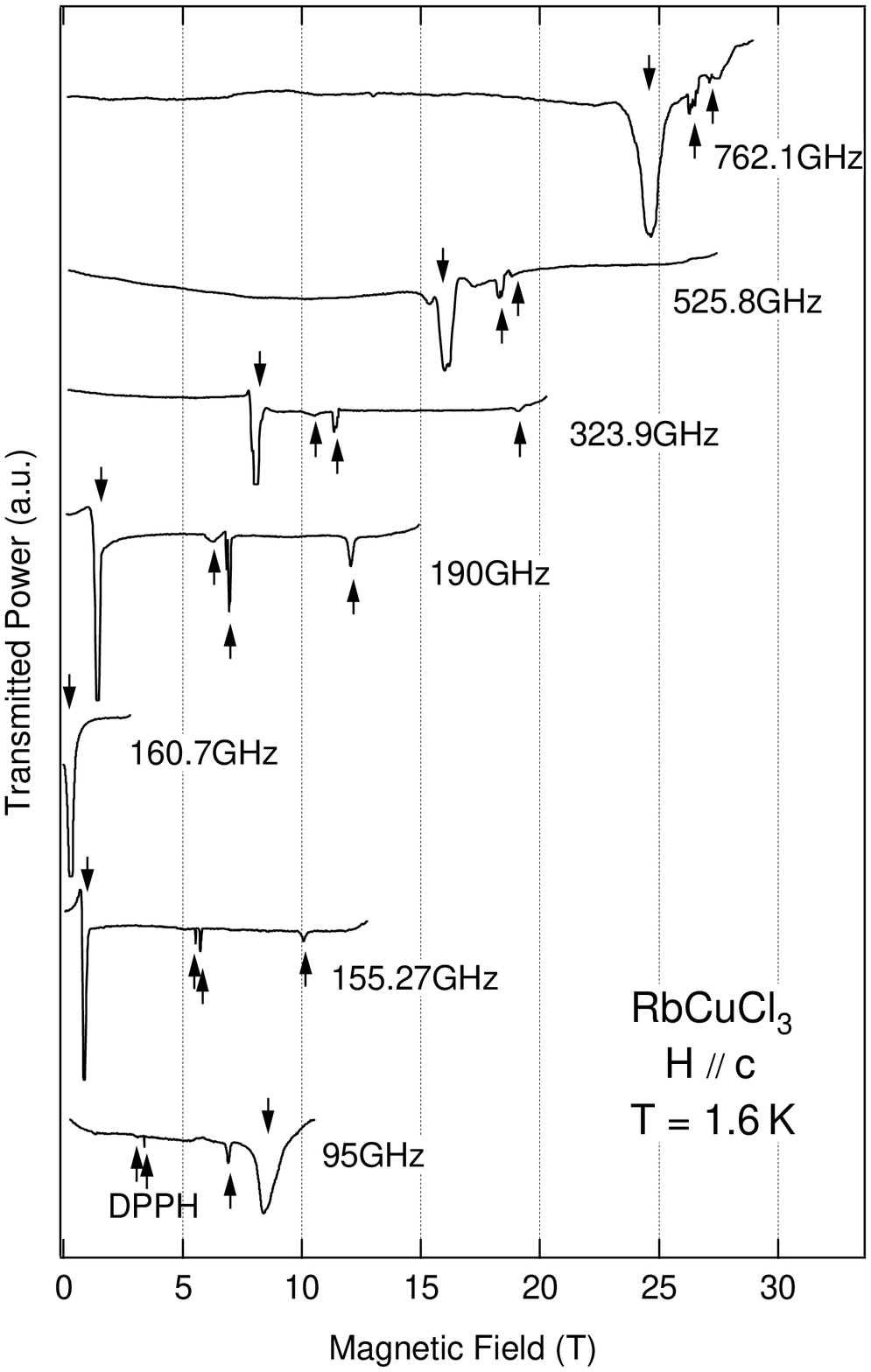,width=80mm,height=120mm}
        (a)
    \end{center}
  \end{minipage}
  \begin{minipage}{90mm}
    \begin{center}
      \epsfigure{file=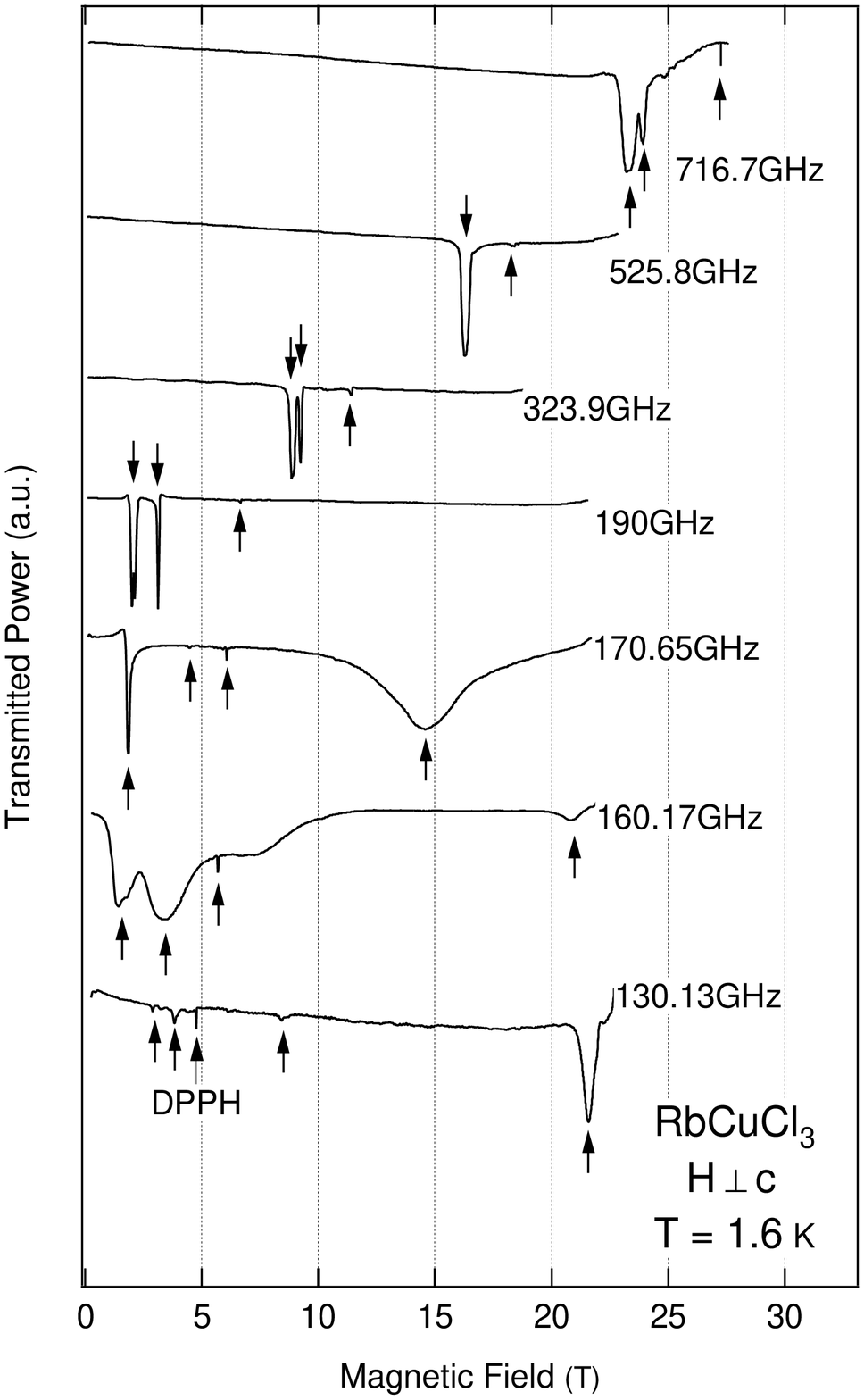,width=80mm,height=120mm}
        (b)
    \end{center}
  \end{minipage}
  \caption{ESR absorption spectra of RbCuCl$_{3}$ observed at 1.6 K for (a) $H{\parallel}c$ and (b)   $H{\perp}c$. Resonance fields are indicated by arrows.}
  \label{fig:5}
\end{fullfigure}
Figure 5 shows the ESR absorption spectra of RbCuCl$_{3}$ observed at 1.6 K for $H{\parallel}c$ and $H{\perp}c$. The resonance fields are indicated by arrows. In Fig. 6, we summarize the resonance positions obtained at 1.6 K. Strong resonance modes are represented by closed symbols. We see that the frequency vs. field diagrams for strong resonances of RbCuCl$_{3}$ are similar to those of CsCuCl$_{3}$.\cite{Palme,Tanaka,Ohta1,Ohta2,Schmidt}\par
\begin{fullfigure}[htbp]
  \begin{minipage}{90mm}
    \begin{center}
      \epsfigure{file=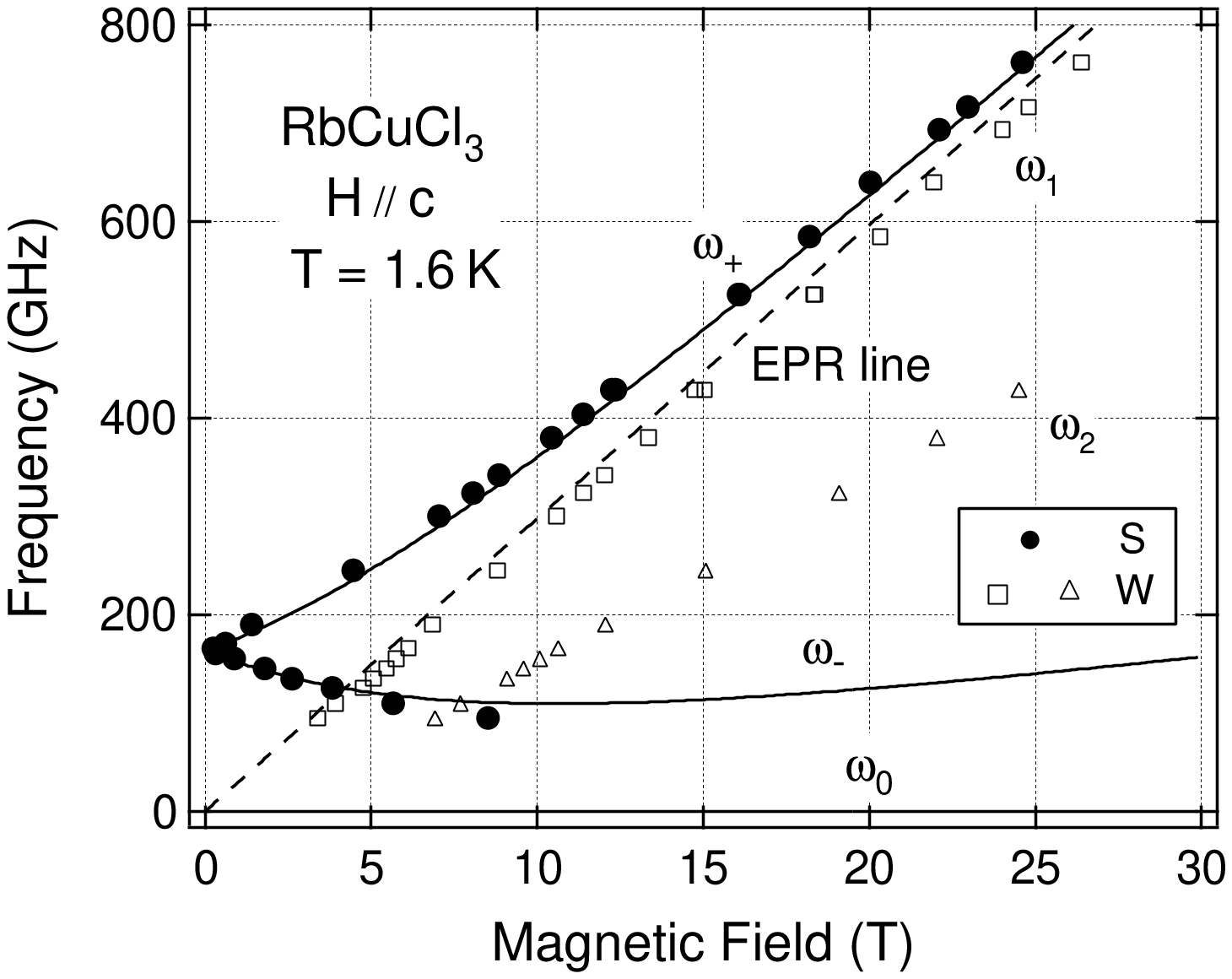,width=80mm}
      \vspace{3mm}
        (a)
    \end{center}
  \end{minipage}
  \begin{minipage}{90mm}
    \begin{center}
      \epsfigure{file=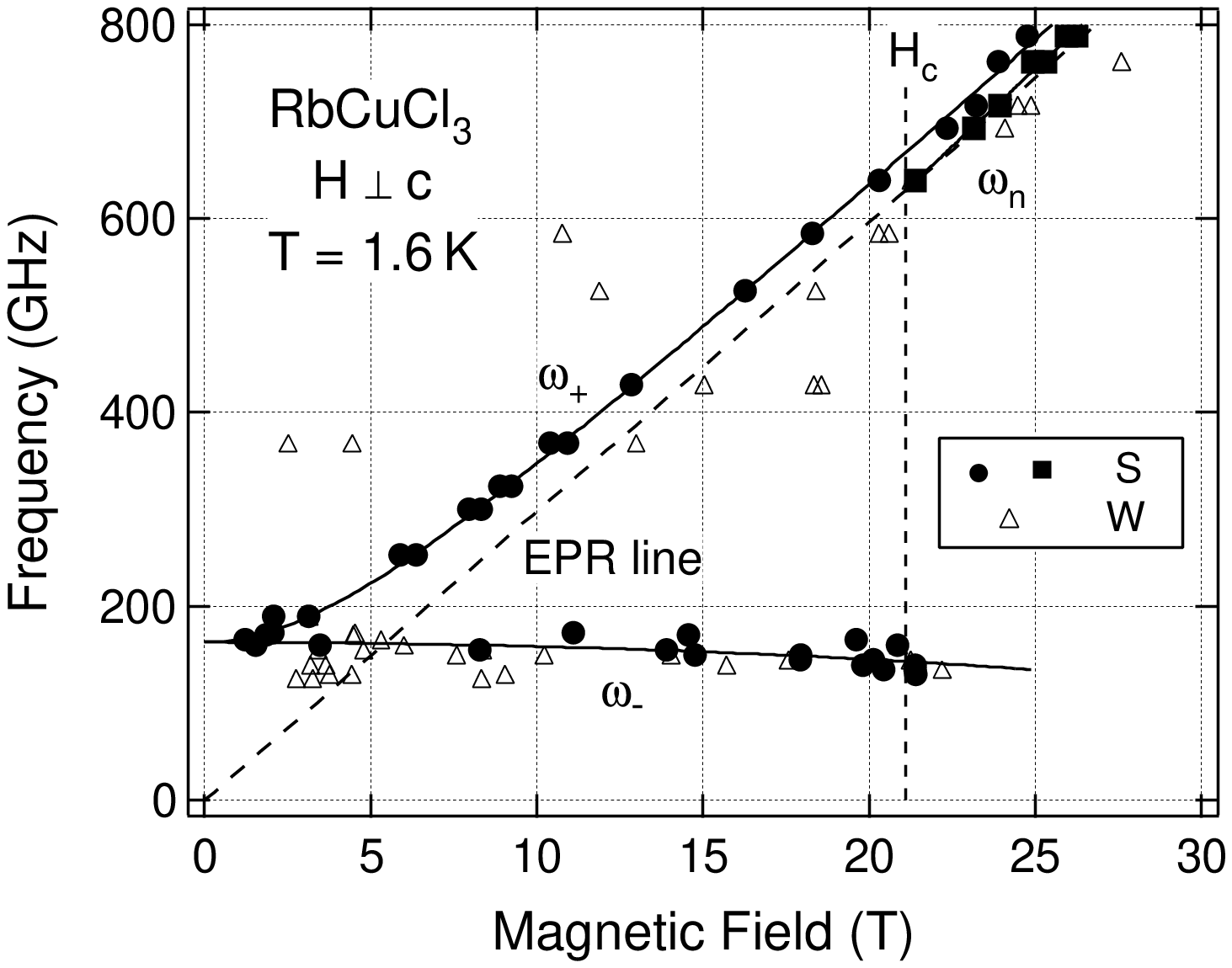,width=80mm}
      \vspace{3mm}
        (b)
    \end{center}
  \end{minipage}
  \caption{Frequency versus field diagrams of RbCuCl$_{3}$ obtained at 1.6 K for (a) $H{\parallel}c$ and (b)   $H{\perp}c$. Closed and open symbols denote strong and weak resonances, respectively. Solid lines in Fig. 6(a)   are fitting curves of eq. (3.6) and those in (b) are the guide for the eyes.}
  \label{fig:6}
\end{fullfigure}
For $H{\parallel}c$, two strong resonances and two weak resonances are observed. We labeled the observed ESR modes as shown in Fig. 6(a). Strong $\omega_+$ and $\omega_-$ modes have almost the same zero-field gap $\Delta=163$ GHz. The frequency of the $\omega_+$ mode increases with the external field, while that of the $\omega_-$ mode decreases with the external field. The resonance frequencies of weak $\omega_1$ and $\omega_2$ modes are almost proportional to the external field, and the $\omega_1$ mode is close to the paramegnetic resonance (EPR) line. \par
When the in-plane anisotropy is negligible, so that all spins are equivalent for $H{\parallel}c$, we can derive the resonance conditions using the formula given by Cooper {\it et al}.\cite{Cooper} For $H{\parallel}c$, we have three observable modes, which are expressed as
\begin{eqnarray}
\hbar{\omega}_{\pm}
& = & S \Bigl(\left\{2J({\mib Q}_0)-J(0)-J(2{\mib Q}_0)\right\} \nonumber \\
& &\hspace{-10mm}\times \bigl(\left\{2J({\mib Q}_0)-J(0)-J(2{\mib Q}_0)\right\}\sin^2{\phi}-4{\Delta}J_0\cos^2{\phi}\bigr) \Bigr)^{1/2} \nonumber \\[2mm]
& &{}\hspace{12mm}\pm\left\{J(2{\mib Q}_0)-J(0) \right\}S\sin{\phi} , 
\end{eqnarray}
and ${\omega}_0=0$. Since spins are arranged ferromagnetically along the $c$-axis, the intrachain exchange interaction $J_0$ does not participate in the resonance conditions. Substituting the interchain exchange interactions $J_1/k_{\rm B}=-10.6$ K and $J_1'/k_{\rm B}=-17.4$ K, $g_{\parallel}=2.13$ and $Q_0=0.2993$ into eq. (3.6), we obtain ${\Delta}J_0/k_{\rm B}=-0.45$ K. Solid lines in Fig. 6(a) are fitting curves of eq. (3.6). We see that strong $\omega_+$ and $\omega_-$ modes are well described by eq. (3.6). \par
Within the above analysis, we cannot derive the weak $\omega_1$ and $\omega_2$ modes. Since the crystal structure of RbCuCl$_3$ is monoclinic below $T_{\rm t2}=260$ K, a small in-plane anisotropy can exist. Thus, strictly speaking, all spins are not equivalent. We infer that the slight deformation of the ideal helical spin structure makes the weak $\omega_1$ and $\omega_2$ modes detectable. However, the in-plane anisotropy must be much smaller than the planar anisotropy ${\Delta}J_0$, because the splitting of the zero-field gaps for the $\omega_+$ and $\omega_-$ modes is very small. \par
When the external field is perpendicular to the $c$-axis, two strong resonances (${\omega}_+$ and ${\omega}_-$) and several weak resonances are observed for $H<H_{\rm c}$. Each strong resonance splits into two or three resonances. Since the low-temperature crystal structure is monoclinic, there are six kinds of domains. For $H\parallel c$, all of the domains are equivalent, while for $H\perp c$, they are not. We deduce that the splitting of the strong resonance for $H\perp c$ is due to the inequivalent monoclinic domains. For $H>H_{\rm c}$, an additional strong resonance ($\omega_{\rm n}$) appears almost on the EPR line. It is understood that the new $\omega_{\rm n}$ mode is characteristic of the high-field phase, although the spin structure of the phase is unknown. Unfortunately, it is difficult to derive the resonance conditions for $H{\perp}c$, because all spins are not equivalent. \par 

\subsection{Discussion}
It is difficult to distinguish the signs of the exchange interactions only from the susceptibility data, because the magnetic susceptibility of RbCuCl$_3$ can also be described in terms of antiferromagnetic intrachain coupling and ferromagnetic interchain coupling. Therefore, it is natural that Tazuke {\it et al.}\cite{Tazuke1} misjudged the signs of the exchange interactions. From the neutron diffraction experiment,\cite{Reehuis} it is evident that intrachain and interchain exchange interactions are ferromagnetic and antiferromagnetic, respectively. The interchain interactions $J_1$ and $J'_1$ can be determined from the saturation field $H_{\rm s}$ and the ordering vector $\mib Q_0$. Thus, when we evaluated the intrachain interaction $J_0$ from the susceptibility data, we assumed that $J_0>0$ and fixed the values of $J_1$ and $J'_1$. \par
\begin{table}
\caption{Magnetic parameters of RbCuCl$_3$ and CsCuCl$_3$. $D$ is the magnitude of the $\mib D$ vector for the Dzyaloshinsky-Moriya interaction.}
\label{table:1}
\begin{tabular}{@{\hspace{\tabcolsep}\extracolsep{\fill}}ccccc} \hline
 & RbCuCl$_3$  & CsCuCl$_3$  \\ \hline
$T_{\rm N}$	& $18.91$ K, $18.87$ K$\,$\cite{Perez2} & 10.6 K$\,$\cite{Adachi,Weber,Hyodo}  \\
$J_0/k_{\rm B}$	& 25.7 K & 28 K$\,$\cite{Tazuke2,Hyodo} \\
$J_1/k_{\rm B}$	& $-10.6$ K	& $-4.9$ K$\,$\cite{Tanaka}  \\
$J'_1/k_{\rm B}$	& $-17.4$ K	& $-4.9$ K$\,$\cite{Tanaka}  \\
$D/k_{\rm B}$	& 	& 5.0 K$\,$\cite{Adachi}  \\
$\Delta J_0/k_{\rm B}$	& $-0.45$ K	& $-0.24$ K$\,$\cite{Tanaka}  \\ \hline
\end{tabular}
\end{table}
In Table I we summarize the magnetic parameters of RbCuCl$_3$ together with those for CsCuCl$_3$. The intrachain exchange interaction of RbCuCl$_3$ is almost the same as that of CsCuCl$_3$, while the interchain exchange interaction of RbCuCl$_3$ is 2.6 times that of CsCuCl$_3$ on average. The latter is responsible for the fact that the N\'{e}el temperature of RbCuCl$_3$ is about twice that of CsCuCl$_3$. Although in RbCuCl$_3$ interchain interactions $J_1$ and $J'_1$ are smaller than intrachain interaction $J_0$, they are of the same order. Therefore, RbCuCl$_3$ is a three-dimensional spin system rather than a quasi-one-dimensional one. \par
When the external field is perpendicular to the $c$-axis, a small magnetization jump indicative of a phase transition of the first order occurs at $H_{\rm c}=21.2$ T $\approx (1/3)H_{\rm s}$, while for $H\parallel c$, no anomaly is seen in the magnetization curve, as shown in Fig. 4. The behavior of the magnetization process of RbCuCl$_3$ contrasts that of CsCuCl$_3$. In CsCuCl$_3$, a small jump occurs at $H_{\rm c}\sim (1/3)H_{\rm s}$ for $H\parallel c$,\cite{Nojiri1} and a plateau-like anomaly is observed for $H\perp c$.\cite{Nojiri1,Nojiri2} The quantum fluctuation gives rise to the field-induced phase transitions in CsCuCl$_3$.\cite{Nikuni1,Schotte,Nikuni2} \par
RbCuCl$_3$ exhibits two magnetic phase transitions at $T_{\rm N1}=18.91$ K and $T_{\rm N2}=18.87$ K at zero field.\cite{Perez2} With increasing external field perpendicular to the $c$-axis, both transition temperatures decrease, and the temperature range of the intermediate phase between $T_{\rm N1}$ and $T_{\rm N2}$ increases.\cite{Perez2} Therefore, we infer that the phase transition at $H_{\rm c}$ for $H\perp c$ is related to the phase boundary for $T_{\rm N2}$. \par
    Using the Landau theory, Zhitomirsky has investigated the magnetic phase diagram for $J_1 \ne J_1'$ in the case of the antiferromagnetic intrachain interaction $J_0$.\cite{Zhitomirsky} However, his results are not applicable to RbCuCl$_3$ with the ferromagnetic $J_0$, because the field-induced phase transitions for $J_0 > 0$ and $J_0 < 0 $ are quite different. \par
The classical molecular field theory\cite{Nagamiya1,Nagamiya2} predicts that a transition from a helical spin structure to a fan structure occurs, when an external field is applied in the easy-plane. Thus, it is possible to attribute the field-induced phase transition at $H_{\rm c}=21.2$ T for $H\perp c$ to the {\it helix-fan} transition. The critical field $H_{\rm c}$ for the {\it helix-fan} transition depends on the ordering vector $\mib Q_0$ for the helical structure. The molecular field theory predicts that $H_{\rm c}>(\sqrt 2-1)H_{\rm s}$.\cite{Nagamiya1,Nagamiya2} In RbCuCl$_3$, the transition field is $H_{\rm c}\approx (1/3)H_{\rm s}$, which is lower than the lower limit $(\sqrt 2-1)H_{\rm s}$ for the {\it helix-fan} transition. Thus at the moment, it too early to conclude that the transition at $H_{\rm c}=21.2$ T corresponds to the {\it helix-fan} transition. The neutron diffraction experiment for $H>H_{\rm c}$ is necessary to elucidate the nature of the field-induced phase transition in RbCuCl$_3$.

\section{Conclusion}
We have presented the results of magnetic susceptibilities, high-field magnetization processes and ESR measurements for the $S=1/2$ stacked triangular antiferromagnet RbCuCl$_3$ with helical spin ordering in the basal plane. From our analyses, the magnetic interactions in RbCuCl$_3$ were determined, as shown in Table I. For $H{\perp}c$, a phase transition of the first order was observed at $H_{\rm c}=21.2$ T, the value of which is almost one-third of the saturation field. The nature of the field-induced phase transition, however, remains. The strong ESR modes for $H\parallel c$ were well described in terms of the ideal helical spin structure. This indicates that the in-plane anisotropy is negligible, although the crystal lattice is monoclinic.

\section*{Acknowledgements}
The authors would like to acknowledge M. Reehuis and U. Schotte for stimulating discussions. They also express their sincere thanks to T. Ono for useful discussion and technical support. Part of this work was carried out under the Visiting Researcher's Program of the Institute of Materials Research, Tohoku University.


\begin{thebibliography}{99}
\bibitem{Collins} M. F. Collins and O. A. Petrenko, Can. J. Phys. {\bf 75} (1997) 605 and references therein.
\bibitem{Nojiri1} H. Nojiri, Y. Tokunaga and M. Motokawa: 
J. Phys. (Paris) {\bf 49} (1988) Suppl. C8 p. 1459.
\bibitem{Nikuni1} T. Nikuni and H. Shiba: 
J. Phys. Soc. Jpn. {\bf 62} (1993) 3268.
\bibitem{Schotte} U. Schotte, N. St\"{u}\ss er, K. D. Schotte, H. Weinfurter, H. M. Mayer and M. Winkelmann: 
J. Phys.: Condens. Matter {\bf 6} (1994) 10105.
\bibitem{Ono1} T. Ono, H. Horai and H. Tanaka: 
J. Phys.: Condens. Matter {\bf 12} (2000) 975.
\bibitem{Ono2} T. Ono, H. Tanaka, T. Kato, A. Hoser, U. Schotte and N. St\"{u}\ss er: 
preprint.
\bibitem{Crama} W. J. Crama: 
J. Solid State Chem. {\bf 39} (1981) 168.
\bibitem{Harada} M Harada: 
J. Phys. Soc. Jpn. {\bf 52} (1983) 1646.
\bibitem{Tazuke1} Y. Tazuke, S. Kinouchi, H. Tanaka, K. Iio and K. Nagata: 
J. Phys. Soc. Jpn. {\bf 55} (1986) 4020.
\bibitem{Reehuis} M. Reehuis, R. Feyerherm, U. Schotte, M. Meschke and H. Tanaka: 
submitted to J. Phys. Chem. Solids.
\bibitem{Glinka} C. J. Glinka, V. J. Minkiewicz, D. E. Cox and C. P. Khattak: 
J. Magn. Magn. Mater. {\bf 18} (1973) 659.
\bibitem{Kato} T. Kato, K. Iio, T. Hoshino, T. Mitsui and H. Tanaka: 
J. Phys. Soc. Jpn. {\bf 61} (1992) 275.
\bibitem{Perez1} F. P\'{e}rez, T. Werner, J. Wosnitza, H. v. L\"{o}hneysen and H. Tanaka: 
Phys. Rev. B {\bf 58} (1998) 9316.
\bibitem{Perez2} F. P\'{e}rez-Willard, A. Fai\ss t, J. Wosnitza. H. v. L\"{o}hneysen, U. Schotte and H. Tanaka: 
Eur. Phys. J. B {\bf 15} (2000) 247.
\bibitem{Adachi} K. Adachi, N. Achiwa and M. Mekata: 
J. Phys. Soc. Jpn. {\bf 49} (1980) 545.
\bibitem{Selwood} P. W. Selwood: 
{\it Magnetochemistry} (Interscience, New York, 1956) 2nd ed., Chap. 2, p. 78.
\bibitem{Rushbrooke1} G. S. Rushbrooke and P. J. Wood: 
Mol. Phys. {\bf 1} (1958) 257.
\bibitem{Rushbrooke2} G. S. Rushbrooke, G. A. Baker  and P. J. Wood: 
{\it Phase Transitions and Critical Phenomena} (Academic Press, London, 1974) Vol. 1, p. 246.
\bibitem{Tazuke2} Y. Tazuke, H. Tanaka, K. Iio and K. Nagata: 
J. Phys. Soc. Jpn. {\bf 50} (1981) 3919.
\bibitem{Palme} W. Palme, F. Mertens, O. Born, B. L\"{u}thi and U. Schotte: 
Solid State Commun. {\bf 76} (1990) 873.
\bibitem{Tanaka} H. Tanaka, U. Schotte and K. D. Schotte: 
J. Phys. Soc. Jpn. {\bf 61} (1992) 1344.
\bibitem{Ohta1} H. Ohta, S. Imagawa, M. Motokawa and H. Tanaka: 
J. Phys. Soc. Jpn. {\bf 62} (1993) 3011.
\bibitem{Ohta2} H. Ohta, S. Imagawa, M. Motokawa and H. Tanaka: 
Physica B {\bf 201} (1994) 208.
\bibitem{Schmidt} S. Schmidt, B. Wolf, M. Sieling, S. Zvyagin, I. Kouroudis and B. L\"{u}thi: 
Solid State Commun. {\bf 108} (1998) 509.
\bibitem{Cooper} B. R. Cooper, R. J. Elliot, S. J. Nettel and H. Suhl: Phys. Rev. {\bf 127} (1962) 57.
\bibitem{Weber} H. B. Weber, T. Werner, J. Wosnitza, H. v. L\"{o}hneysen and U. Schotte: 
Phys. Rev. B {\bf 54} (1996) 15924.
\bibitem{Hyodo} H. Hyodo, K. Iio and K. Nagata: 
J. Phys. Soc. Jpn. {\bf 50} (1981) 1545.
\bibitem{Nojiri2} H. Nojiri, K. Takahashi, T. Fukuda, M. Fujita, M. Arai and M. Motokawa: 
Physica B {\bf 241-243} (1998) 210.
\bibitem{Nikuni2} T. Nikuni and A. E. Jacobs: 
Phys. Rev. B {\bf 57} (1998) 5205.
\bibitem{Zhitomirsky} M. E. Zhitomirsky: Phys. Rev. B {\bf 54} (1996) 353.
\bibitem{Nagamiya1} T. Nagamiya, K. Nagata and Y. Kitano: Prog. Theor. Phys. {\bf 27} (1962) 1253.
\bibitem{Nagamiya2} T. Nagamiya: Solid State Phys. {\bf 20} (1967) 305.

\end{thebibliography}
\end{document}